\documentclass[twocolumn,showpacs,preprintnumbers,amsmath,amssymb]{revtex4}
\usepackage{graphicx}
\usepackage{dcolumn}
\usepackage{bm}

\begin{document}

\title{Plasmon and coupled plasmon-phonon modes in graphene}

\author{H.M. Dong$^{1,2}$}
\author{S.H. Zhang$^{1}$}
\author{W. Xu$^{1,3}$}\email{wenxu_issp@yahoo.cn}
\address{$^1$Key Laboratory of Materials Physics, Institute of Solid State Physics,
Chinese Academy of Sciences, China}
\address{$^2$Max-Planck Institute for the Science of Light, 91058 Erlangen, Germany}
\address{$^3$Department of Physics, Yunnan University, Kunming 650091,
China}

\date{\today}

\begin{abstract}

Plasmon and coupled plasmon-phonon modes in graphene are
investigated theoretically within the diagrammatic self-consistent
field theory. It shows that two plasmon modes and four coupled
plasmon-phonon modes can be excited via intra- and inter-band
transition channels. It is found that with increasing $q$ and
carrier density, the plasmon modes couple strongly with the
optic-phonon modes in graphene. The coupled plasmon-phonon modes
exhibit some interesting features which can be utilized to realize
the plasmonic devices. Our results suggest that the carrier-phonon
interaction should be considered to understand and explain the
properties of elementary electronic excitations in graphene.

\end{abstract}

 \maketitle

\section{Introduction}
Since graphene was discovered in 2004 \cite{ksn}, it has brought us
many surprises and opened up a new field of research in condensed
matter physics and nano-electronics. From a physics point of view,
carriers in graphene are massless, gapless and relativistic Dirac
particles with nearly linear energy dispersion. Thus graphene brings
us a great opportunity to look into the many-body interactions in a
Dirac quasi-particle system. Recently the elementary electronic
excitation via plasmon modes resulting from Coulomb interaction has
been intensively investigated for graphene based electronic systems.
Das Sarma and Hwang \cite{das} and Wang and Chakraborty \cite{wxf}
have found theoretically that the plasmon frequency of the Dirac
fermions in graphene shows very different behaviors from the usual
2DEG systems. More interestingly, it has been shown that the plasmon
frequencies in graphene can be within the terahertz ($10^{12}$ Hz or
THz) bandwidth \cite{vra} and 1 - 10 THz sources based on plasmon
amplification in graphene can be generated for moderate carrier
densities by applying the gate voltages \cite{fr}. Moreover, Jablan
\textit{et.al.} has demonstrated that the plasmon excitation in
graphene can have low losses for infrared frequencies \cite{jabl}
and proposed that such a feature can be utilized for nano-photonic
device applications. The results obtained from further theoretical
investigations have indicated that the guided plasmon waves in
graphene p-n junctions can be achieved and graphene can then be
applied as a material for nano-plasmonics \cite{egm}. However, here
we notice that most of the published theoretical studies mentioned
above were carried out without considering the carrier-phonon
coupling in graphene and only the intra-band transition channel was
taken into consideration. Very recently, Bostwick \textit{et.al.}
observed the plasmarons in free-standing doped graphene, which
results from bound states of charge carriers with plasmons
\cite{plaron}.

On the other hand, recent experimental work has suggested that the
adiabatic Born-Oppenheimer approximation fails in graphene because
of the presence of the interactions between the peculiar Dirac
fermions and optic phonons \cite{pisa}. It has been experimentally
demonstrated that the plasmon modes in epitaxial graphene on SiC
wafer can be strongly coupled with the surface optical phonon modes
in silicon carbide \cite{liuy}. Bostwick \textit{et.al.} have found
experimentally that the electron-electron interaction and the
electron-phonon coupling should be considered on an equal footing to
understand the properties of Dirac quasi-particles in graphene
\cite{bos}. Siegel \textit{et.al.} have experimentally demonstrated
electron-electron interaction results in unique renormalizations,
electron-phonon interaction diminish with decreasing doping in
graphene \cite{pnas}. The measurement of the Raman spectrum has now
been widely used to investigate the properties of photo-excited
electronic excitations in graphene systems \cite{cst,dmb}. The
results obtained from recent Raman measurements have shown that the
position of optic-phonon$-$induced G-peak depends very weakly on
carrier density in the sample. However, the Raman intensity around
the G-peak increases with increasing doping concentration in
graphene. Such an effect is an identification that both the
electron-electron interaction and the electron-phonon coupling
affect strongly the elementary electronic excitations in graphene.
These experimental findings have clearly indicated that the
carrier-phonon coupling is very important in understanding and
explaining the electronic and optoelectronic properties of graphene,
especially the elementary electronic excitations.

We know that in conventional electronic gas systems the plasmon
modes can be strongly modified by phonon scattering so that the new
modes, namely the coupled plasmon-phonon modes, are formed. Raman
scattering \cite{moo}, infrared and photoemission measurements
\cite{tak}, and ultrafast pump-and-probe experiments \cite{alt},
{\it etc.} have been widely applied to detect and study the coupled
plasmon-phonon modes in semiconductor based bulk and low-dimensional
electronic systems. The investigation of the plasmon and coupled
plasmon-phonon excitations from electron gas systems is also a
fundamental basis for the realization of these systems as plasmonic
devices for various applications. Because graphene is an ideal 2DEG
system with high carrier density and high carrier mobility, it is
natural for us to expect that graphene can exhibit more excellent
and unique plasmonic properties for device applications. Thus, in
order to gain an in-depth understanding of the many-body
interactions in graphene systems and, on this basis, to explore the
potential applications of graphene as practical nano-plasmonic
devices, it is necessary and significant to examine the roles which
the carrier-carrier interaction and the carrier-phonon coupling can
play in such a Dirac fermion system. And this becomes the prime
motivation of the present theoretical investigation.

\section{Theoretical approach}

Here we consider a graphene sheet in the $xy$-plane. A carrier
(electron or hole) in a monolayer graphene can be described by the
Dirac equation for a massless neutrino. The energy spectrum and
wavefunction for a Dirac quasi-particle in graphene in the absence
of the scattering mechanisms can be obtained analytically. They are,
respectively, $E_\lambda({\bf k})=\lambda\gamma|{\bf
k}|=\lambda\gamma k$ and $\psi_{\lambda{\bf k}}({\bf r})=|{\bf
k},\lambda>=2^{-1/2}[1, \lambda e^{i\phi} ] e^{i{\bf k}\cdot {\bf
r}}$ in the form of a row matrix. Here, ${\bf k}$ is the wavevector
for a carrier and $k=\sqrt{k_x^2+k_y^2}$, ${\bf r}=(x,y)$,
$\gamma=\hbar v_F $ is the band parameter with $v_F\simeq 10^8$ cm/s
being the Fermi-velocity of a Dirac quasi-particle, $\lambda=+1$ for
an electron and $\lambda=-1$ for a hole, and $\phi$ is the angle
between ${\bf k}$ and the $x$ direction.

With the single-particle wavefunction for a carrier, we can
calculate the electrostatic energy induced by bare carrier-carrier
(c-c) interaction in graphene through
\begin{eqnarray}\label{a3}
V_{\lambda_1'\lambda_1\lambda_2'\lambda_2}^{cc}&=&S_{\lambda_1'\lambda_1
\lambda_2'\lambda_2} \int d^2 {\bf r}_1 d^2 {\bf r}_2\
\psi_{\lambda_1'{\bf
k}_1'}^*({\bf r}_1)\psi_{\lambda_1{\bf k}_1}({\bf r}_1)\nonumber\\
 & &\times V({\bf r}_1-{\bf r}_2)\psi_{\lambda_2'{\bf k}_2'}^*({\bf r}_2)
 \psi_{\lambda_2{\bf k}_2}({\bf
 r}_2),\nonumber\\
\end{eqnarray}
where $V({\bf r})=e^2/\epsilon_\infty |{\bf r}|$ is the Coulomb
potential with $\epsilon_\infty\simeq 1$ being the high-frequency
dielectric constant for graphene layer \cite{wxf}, and
$S_{\lambda_1\lambda_2\lambda_3\lambda_4}=-i^{(\lambda_1+\lambda_2+\lambda_3+
\lambda_4)/2}=\pm 1$ or $\pm i$ is a sign function related to the
charge sign of a carrier in different bands. After considering a
momentum conservation law, namely the momentum that flows into the
c-c scattering conserves with what flows out for different
scattering processes, the bare c-c interaction becomes
\begin{equation}\label{a4}
V_{\alpha\beta}^{cc}=S_{\alpha\beta} V_q G_{\alpha\beta}({\bf
k},{\bf q}),
\end{equation}
where we have defined $\alpha=(\lambda'\lambda)$ for intra- (i.e.,
$\lambda'=\lambda$) and inter-band (i.e., $\lambda'\neq \lambda$)
transition, ${\bf q}=(q_x,q_y)$ is the change of the carrier
wavevector during a c-c scattering event, $V_q=2\pi e^2
/\epsilon_\infty q$ is the two-dimensional Fourier transform of the
Coulomb potentail, and
$$G_{\alpha\beta}({\bf k},{\bf q})={1+\alpha A_{\bf kq}\over
2}\delta_{\alpha,\beta}+i{\alpha B_{\bf kq}\over
2}(1-\delta_{\alpha,\beta}),$$ with $A_{\bf kq}=(k+q{\rm
cos}\theta)/|{\bf k}+{\bf q}|$, $B_{\bf kq}=(q{\rm sin}\theta)/|{\bf
k}+{\bf q}|$ and $\theta$ being the angle between ${\bf k}$ and
${\bf q}$.

In the present study, we assume that the graphene system can be
separated into the carriers of interest and the rest of graphene
crystal. For the case of carrier interactions with 2D-like phonons,
the interaction Hamiltonian takes a form
\begin{equation}\label{a5} H_{c-p}=W_{\bf q}a_{\bf q}e^{i({\bf
q}\cdot {\bf r}+\omega_q t)} +W_{\bf q}^* a_{\bf q}^\dagger
e^{-i({\bf q}\cdot{\bf r}+\omega_q t)},
\end{equation}
where ${\bf q}=(q_x,q_y)$ is the phonon wavevector along the
xy-plane, $(a_{\bf q}^\dagger, a_{\bf q})$ are the canonical
conjugate coordinates of the phonon system, $W_{\bf q}$ is the
carrier-phonon (c-p) interaction coefficient, and $\omega_q$ is the
phonon frequency. The Fourier transform of the matrix element for
bare c-p interaction can be written as
\begin{equation}\label{a6}
W_{\alpha}^{cp}=D_0(\omega_q,\Omega)|U_\alpha ({\bf k}, {\bf q})|^2,
\end{equation}
where $\Omega$ is the electronic excitation frequency,
$D_0(\omega_q,\Omega)=2\hbar\omega_q/[(\hbar\Omega)^2-(\hbar\omega_q)^2]$
is the bare phonon propagator, and $|U_{\lambda'\lambda} ({\bf k},
{\bf q})|^2=|<{\bf k+q},\lambda'|W_{\bf q}|{\bf k},\lambda>|^2$. The
published experimental results have indicated that the optic-phonon
coupling is a major factor in determining the intensity and
peak-position of phonon-related Raman spectrum in graphene
\cite{cst,dmb}. Thus, in conjunction with these experimental
findings, in this study we consider carrier interaction with only
optic-phonons. On the basis of a valence-force-field model, the
coupling coefficient for carrier interactions with long-wavelength
optic-phonons in graphene is \cite{tse}
\begin{equation}\label{b1}
W_{\bf q}^\mu=-g M_{\bf q}^\mu, \end{equation} where $g=\hbar\gamma
( B/b^2) /\sqrt{2\rho\hbar\omega_0} $ with $\rho\simeq 6.5\times
10^{-8}$ g/cm$^2$ being the areal density of the graphene sheet,
$\omega_0=196$ meV the optic-phonon frequency at the $\Gamma$-point,
$B\sim 2$ is a dimensionless parameter, and $b=a/\sqrt{3}$ is the
equilibrium bond length. Furthermore,
\begin{equation}\label{b2}
M_{\bf q}^l= \left[
\begin{array}{cc}
  0 & -e^{-i\phi_q}\\
  e^{i\phi_q} & 0 \\
\end{array}\right] \  {\rm and} \
M_{\bf q}^t= \left[
\begin{array}{cc}
  0 & ie^{-i\phi_q}\\
  ie^{i\phi_q} & 0 \\
\end{array}\right],
\end{equation}
for coupling with, respectively, the longitudinal ($l$) and
transverse ($t$) phonon modes, where $\phi_q$ is the angle between
${\bf q}$ and the x-axis. Thus, the squares of the carrier-phonon
scattering matrix elements are
$$|U_{\lambda\lambda'}^{l}({\bf q},{\bf k})|^2=(g^2/2)[1-\lambda'\lambda
{\rm cos}(\phi+\phi'-2\phi_q)]$$ and
$$|U_{\lambda\lambda'}^{t}({\bf q},{\bf k})|^2=(g^2/2)[1+\lambda'\lambda
{\rm cos}(\phi+\phi'-2\phi_q)].$$ The total contribution of the
optic-phonon scattering is
\begin{equation}
|U_{\lambda\lambda'}({\bf q},{\bf k})|^2= |U_{\lambda\lambda'}^{
l}({\bf q},{\bf k})|^2+ |U_{\lambda\lambda'}^{t}({\bf q},{\bf k})|^2
=g^2\gamma^2,
\end{equation}
which is independent on ${\bf q}$ and ${\bf k}$. This feature is
very distinct from that in the conventional semiconductor-based 2DEG
systems.

In a diagrammatic self-consistent field theory \cite{wxu}, the
effective c-c interaction in the presence of the c-p coupling is
given by
$$V_{\alpha\beta}^{eff}=[V_{\alpha\beta}^{cc}+W_{\alpha}^{cp}]
\epsilon_{\alpha\beta}^{-1},
$$
where $\epsilon_{\alpha\beta}=\delta_{\alpha,\beta}\delta({\bf
k})-[V_{\alpha\beta}^{cc}+W_{\alpha\beta}^{cp}]\Pi_\beta ({\bf
k},{\bf q};\Omega)$ is the dynamical dielectric function matrix
element and
$$
\Pi_{\lambda'\lambda}({\bf k},{\bf q};\Omega)=g_sg_v
{f_{\lambda'}[E_{\lambda'}({\bf k}+{\bf q})]- f_\lambda
[E_\lambda({\bf k})]\over \hbar\Omega +E_{\lambda'}({\bf k}+{\bf
q})-E_\lambda ({\bf k})+i\delta}
$$is the pair bubble or density-density correlation function in the
absence of the c-c screening, with $f_\lambda(x)$ being the
Fermi-Dirac function for a carrier in the $\lambda$ band and $g_s=2$
and $g_v=2$ counting respectively the spin and valley degeneracy.
After summing $\epsilon_{\alpha\beta}$ over ${\bf k}$ and setting
$j=(\lambda'\lambda)=1=(++),\ 2=(+-),\ 3=(-+)$ and $4=(--)$, the
dielectric function matrix is obtained as
\begin{equation}\label{a7}
\epsilon(q,\Omega)= $$$$\left[
\begin{array}{cccc}
  1+a_1+b_1 & 0 & 0 & -a_4\\
  0 & 1-a_2-b_2 & -a_3 & 0 \\
  0 & -a_2 & 1-a_3-b_3 & 0 \\
  -a_1 & 0 & 0 & 1+a_4+b_4 \\
\end{array}\right],
\end{equation}
where $$a_{\lambda'\lambda}(q,\Omega)=-(V_q/2)\sum_{\bf k}
(1+\lambda'\lambda A_{\bf kq}) \Pi_j({\bf k},{\bf q}; \Omega)$$ is
induced by the c-c interaction, and $$
b_{\lambda'\lambda}(q,\Omega)=-D_0(\omega_q,\Omega)\sum_{\bf
k}|U_{\lambda'\lambda}({\bf k},{\bf q})|^2 \Pi_j({\bf k},{\bf q};
\Omega)$$ is caused by the c-p coupling. In this study we use a
matrix to present the dielectric function. In contrast to previous
theoretical study in which only the intra-band transition is
considered for calculating the dynamical dielectric function, the
present work includes the contributions from inter-band transition
channels as well. Thus, the dielectric function matrix is a $4\times
4$ matrix. In the absence of the phonon scattering, i.e., when
$b_j=0$, the dynamical dielectric function matrix becomes that
obtained under the standard random phase approximation (RPA). The
determinant of the dielectric function matrix Eq. (\ref{a7}) is
given by
$$|\epsilon|=[(1+a_1+b_1)(1+a_4+b_4)-a_1a_4]$$\vskip-0.5truecm
\begin{equation}\label{a8}
\quad\quad \times [(1-a_2-b_2)(1-a_3-b_3)-a_2a_3],
\end{equation}
which results from intra- (i.e., $j=1$ or $4$) and inter-band (i.e.,
$j=2$ or $3$) electronic transitions. The modes of the elementary
electronic excitation are determined by ${\rm Re}|\epsilon|\to 0$.

Now we consider a n-type (or positively gated) graphene with an
electron density $n_e$ and a distribution function
$f_e(x)=[1+e^{(x-E_F)/k_BT}]^{-1}$ where $E_F$ is the Fermi energy
(or chemical potential) for electrons in the conduction band. In
such a case, the valence band is fully occupied so that the
distribution $f_h(x)=0$, the hole density $n_h=0$, and $a_4=b_4=0$.
The approach can also be applied to the p-type (or negatively gated)
graphene samples because the electrons and holes are symmetrical in
graphene system. At a long-wavelength ($q\to 0$) and low-temperature
($T\to 0$) limit, the frequency of the collective excitation of the
uncoupled plasmon mode ($b_{j}=0$) is obtained as
\begin{equation}\omega_p=\sqrt{2e^2E_Fq/\epsilon_\infty}\sim q^{1/2}
\sim n_e^{1/4},
\end{equation}
which is induced by intra-band transition within the conduction
band. Here, $E_F=\gamma $k$_F$ with k$_F$$=\sqrt{\pi n_e}$ being the
Fermi wavevector in graphene and q is defined as the plasmon
wavevector or momentum. By solving the equation
\begin{equation}\label{a1}
\Omega-{\omega_p^2\over 4E_F}\ln\Big|{2E_F+\Omega\over
2E_F-\Omega}\Big|=0,
\end{equation}
we can obtain the uncoupled plasmon mode $\omega'_p$ induced by
inter-band transition for electrons from valence band to conduction
band in graphene. Similar to a conventional 2DEG system, the
single-particle (electron-hole pair) excitations (SPE) for intra-
and inter-band transition channels are also allowed in graphene. The
intra- (upper) and inter- (lower) band transition boundaries are
still given respectively by $\omega_s=\gamma|\bf {k+q}|\mp
\gamma|k|$. Thus, we are able to identify the intra-band SPE regime
$0 \leq \omega_s \leq \gamma q$ for $q \leq 2$k$_F$, and $\gamma
q-2E_F \leq \omega_s \leq \gamma q$ for $q \geq 2$k$_F$. The
inter-band SPE regime is allowed for $2E_F-\gamma q \leq \omega_s
\leq 2E_F+\gamma q$ in graphene. Furthermore, a critical parameter
$q_c=$k$_F$$(2+c-\sqrt{4c+c^2})$ can be introduced to indicate the
minimum momentum transfer with $c=e^2/(\epsilon_\infty\gamma)$,
where the plasmon mode enters to the electron-hole continuum.

Including the electron-phonon coupling $(b_j\neq0)$, we obtain two
coupled plasmon-phonon modes,
\begin{equation}
\Omega_{\pm}={1\over\sqrt{2}}[\omega_0^2+\omega_{p}^2
\pm\sqrt{(\omega_0^2-\omega_{p}^2)^2+ \omega_0\omega^3_{q}}~]^{1/2}
\end{equation}
induced by intra-band transition for electrons within the conduction
band in graphene, where $\omega_{q}^3=8g^2\gamma^2q^2E_F/\pi$.
Moreover, by solving the equation
\begin{equation}\label{a2}
\Omega-{\pi\omega_p^2(\Omega^2-\omega_0^2)+
4g^2E_F\omega_0\Omega^2\over4\pi
E_F(\Omega^2-\omega_0^2)+16g^2E_F^{2}\omega_0}
\ln\Big|{2E_F+\Omega\over 2E_F-\Omega}\Big|=0,
\end{equation}
we acquire two coupled plasmon-phonon modes $\Omega_{1}$ and
$\Omega_{2}$ induced via inter-band transition for electrons from
valence band to conduction band in graphene system. The analytical
results presented in this Section show that after including the
inter-band transition channels and the electron-phonon coupling, the
new modes of elementary electronic excitations can be achieved in
graphene.

\section{Numerical results and discussions}

\begin{figure}
\centering
\includegraphics[width=0.45\textwidth,angle= 0]{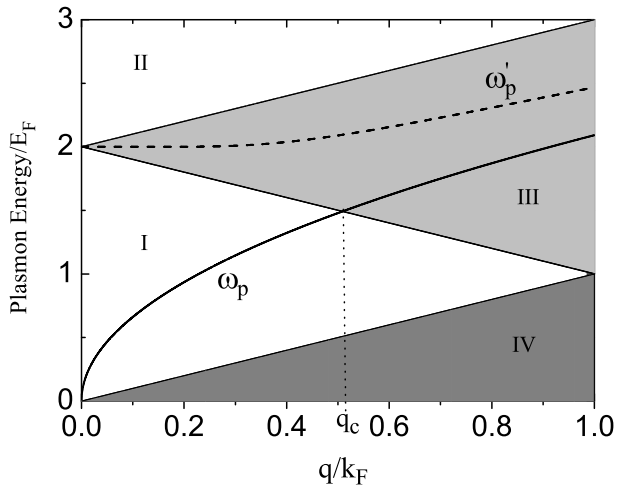}
\caption{Dispersion relation for uncoupled plasmon modes $\omega_p$
(solid line) and $\omega'_p$ (dashed line) at a fixed electron
density $n_e=1 \times 10^{12}$ cm$^{-2}$. The regime III (lightly
shaded) and regime IV (heavily shaded) correspond respectively to
the inter-band SPE and intra-band SPE. The regime I and regime II
are for undamped plasmon excitation and the minimum momentum
transfer $q_c \simeq 0.51$k$_F$ is indicated with
k$_F\simeq1.77\times10^6$ cm$^{-1}$.}
\end{figure}

\begin{figure}
\centering
\includegraphics[width=0.45\textwidth,angle= 0]{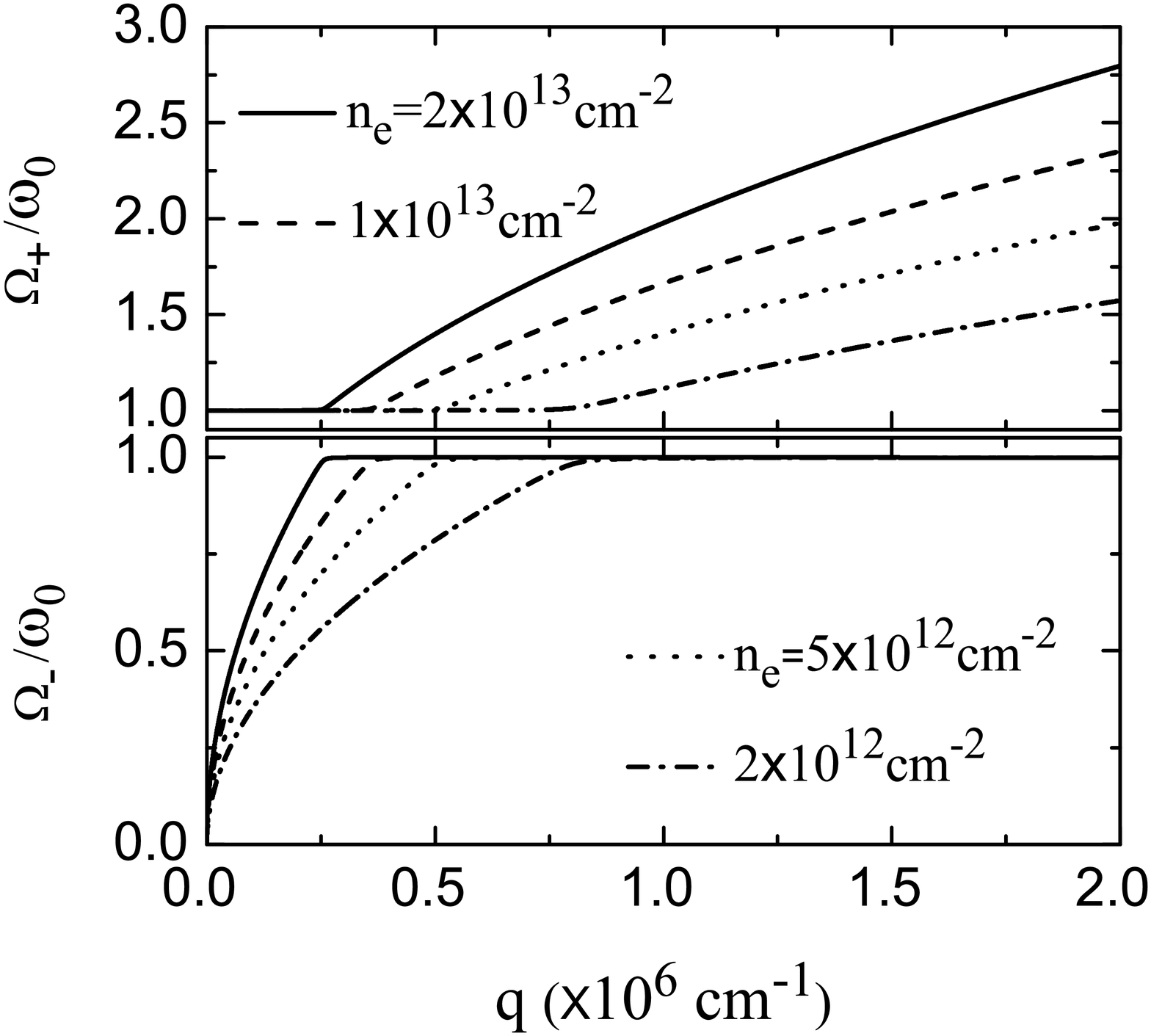}
\caption{Dispersion relation for coupled plasmon-phonon modes
$\Omega_+$ (upper panel) and $\Omega_- $ (lower panel) induced by
intra-band transition for different electron densities in graphene
as indicated. Here $\omega_0=196$ meV is the optic-phonon
frequency.}
\end{figure}

\begin{figure}
\centering
\includegraphics[width=0.45\textwidth,angle= 0]{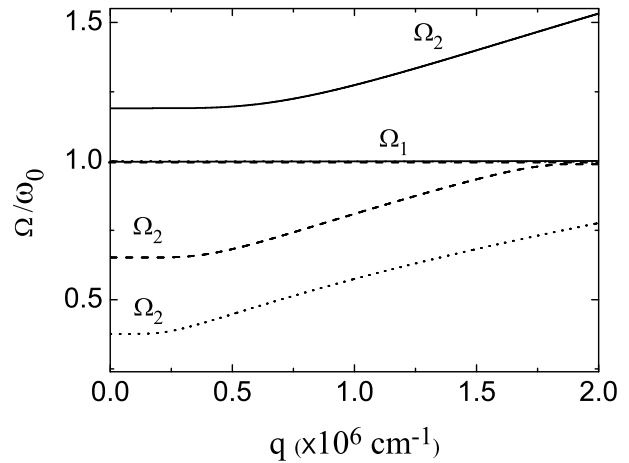}
\caption{Dispersion relation for coupled plasmon-phonon modes
$\Omega_1$ ($\simeq \omega_0$) and $\Omega_2$ ($\geq 2E_F$) induced
by inter-band transitions for different electron densities
$n_e=1\times10^{12}$ cm$^{-2}$ (solid line), $n_e=3\times10^{11}$
cm$^{-2}$ (dashed line) and $n_e=1\times10^{11}$ cm$^{-2}$ (dotted
line). The three curves for $\Omega_1$ coincide roughly.}
\end{figure}

In Fig. 1, the dispersion relation for uncoupled plasmon modes,
$\omega_p$ and $\omega'_p$ induced respectively by intra- and
inter-band transitions, are shown at a fixed electron density in
graphene. As being pointed out by other authors \cite{das,wxf}, the
plasmon frequency induced by intra-band excitation in graphene,
$\omega_p \sim q^{1/2} \sim n_e^{1/4}$, is acoustic-like and depends
strongly on q. The density dependence $\omega_p\sim n_e^{1/4}$ in
graphene differs from that $\omega_p \sim n_e^{1/2}$ in a
conventional 2DEG system. It shows that the plasmon frequency
$\omega_p$ of graphene can be effectively controlled through tuning
carrier densities by gated voltages. In the presence of the
inter-band transition channels, we obtain a new uncoupled plasmon
mode $\omega'_p$ induced by excitation of electrons from occupied
valance band to the empty states in the conduction band. Due to
Pauli exclusion principle, $\omega'_p \geq 2E_F$ is optic-like and
depends relatively weakly on q. With increasing q, $\omega_p$ and
$\omega'_p$ converge in regime III. We find that $\omega_p$ depends
more strongly on carrier density and q than $\omega'_p$ does. For a
undoped graphene sample so that $n_e \to 0$, the plasmon frequency
$\omega_p\to 0$ and the excitation is restricted, whereas there
exits the plasmon mode with a frequency $\omega'_p$ due to
inter-band electronic transition. Thus, this new plasmon mode can be
used to investigate the collective electronic excitation from an
nearly intrinsic graphene. In Fig. 1, we also show the regimes III
and IV for $q \leq $k$_F$ within which the SPE can take place.
Similar to a conventional 2DEG, within the SPE regime the Landau
damping occurs so that the imaginary part of the dielectric function
Im$\epsilon (\omega, q) \neq 0$, and that if the plasmon excitation
exists in these regimes for $q > q_c \simeq 0.51 $k$_F$ (see e.g.
III and IV in Fig. 1), the plasmon modes are strongly damped and
decay into the electron-hole pairs. Thus, one can see that the
plasmon mode $\omega_p$ is undamped for $q < q_c$ in regime I where
Im$\epsilon (\omega, q) = 0$, whereas the plasmon mode $\omega'_p$
is damped in the full regime of III in which Im$\epsilon (\omega, q)
\neq 0$. Consequently, in graphene, the undamped plasmon excitation
can be achieved via intra-band transition, while the plasmon
excitation via inter-band transition is damped due to the Landau
damping effect.

In Fig. 2, the dispersion relation for coupled plasmon-phonon modes
$\Omega_+$ and $\Omega_-$ induced by intra-band transition is shown
for different electron densities. For small values of $q<q_0$, the
lower frequency branch $\Omega_- < \omega_0$ is acoustic-like (or
plasmon-like) which depends strongly on q, whereas the higher
frequency branch $\Omega_+ \simeq \omega_0$ is optic-like (or
phonon-like) which depends rather weakly on q. Interestingly, for
relatively large values of $q>q_0$, the lower frequency branch
becomes optic-like and $\Omega_-\simeq \omega_0$, whereas the higher
frequency branch is acoustic-like and $\Omega_+>\omega_0$. When
$q>q_0$, the plasmon modes couples strongly to the optic-phonon
modes and the frequencies of the coupled plasmon-phonon modes in
graphene differ significantly from the corresponding plasmon and
optic-phonon frequencies. The $q_0$ value, at where the two branches
of the excitations change the nature of the q dependence, decreases
with increasing the electron density. It is interesting to compare
the results for the coupled plasmon-phonon modes in graphene with
those in a conventional 2DEG (C2DEG) system. In a semiconductor
based C2DEG, there exit three modes for coupled plasmon-phonon
excitations induced by intra-subband transition channels via
electron interaction with optic-phonons through Fr\"ohlich coupling
\cite{pee}. They are: i) longitudinal-optic (LO) phonon-like mode
$\omega_+\simeq \omega_{LO}$ with $\omega_{LO}$ being the LO-phonon
frequency; ii) transverse-optic (TO) phonon-like mode
$\omega_-\simeq \omega_{TO}$ with $\omega_{TO}$ being the TO-phonon
frequency; and iii) plasmon-like mode $\omega_1$, respectively. In a
C2DEG, the plasmon-like mode does not coupled markedly with the
phonon-like modes and the phonon-like modes depend quite weakly on q
\cite{pee}. The reason why only two coupled plasmon-phonon modes are
observed in graphene is that at the $\Gamma$-point,
$\omega_{LO}\simeq\omega_{TO}=\omega_0$ for graphene \cite{tse} in
contrast to a C2DEG where normally $\omega_{LO}\neq \omega_{TO}$ at
the $\Gamma$-point. The different q-dependence of the coupled
plasmon-phonon modes in graphene and in a C2DEG is mainly induced by
different energy spectra in two electronic systems. As we know, the
elementary electronic excitation is achieved via varying the energy
and momentum of the electrons. Thus, the energy spectrum of the
electronic system determines mainly the dispersion relation of the
electronic excitation. In contrast to a parabolic energy spectrum
for a C2DEG, graphene has a linear energy spectrum. As a result, the
dispersion relation of the coupled plasmon-phonon modes in graphene
differs significantly from those in a C2DEG. From Fig. 2, we can
also find that with increasing electron density $n_e$, the $q_0$
value decreases, and the plasmon and phonon modes are strongly
coupled. This is because the plasmon frequency $\omega_p \sim
n_e^{1/4}$ in graphene can be comparable to the optic-phonon
frequency $\omega_0$ at a relatively large $n_e$.

In Fig. 3, we show the dispersion relation for coupled
plasmon-phonon modes with frequencies $\Omega_1$ and $\Omega_2$
induced by inter-band transition in graphene for different electron
densities $n_e$. We see that a strong optic-phonon$-$like mode with
a frequency $\Omega_1\simeq \omega_0$ can be generated in graphene
and the frequency of this mode depends very weakly on both q and the
electron density. In contrast, the plasmon-like mode $\Omega_2\sim
\omega'_p$ depends sensitively on electron density and on plasmon
wavevector q. $\Omega_2$ increases with $n_e$ due to its
plasmon-like nature of the excitation. We note that both modes with
$\Omega_1 $ and $\Omega_2$ are optic-like and uncoupled with each
other. For a n-type graphene, an excitation energy $E>E_F$ is
required to excite electrons in the valence band into the conduction
band and to induce the elementary electronic excitations due to the
phase-space restriction. As a result, there are always $\Omega_1
\simeq \omega_0$ and $\Omega_2 \geq 2E_F$ for coupled plasmon-phonon
frequencies induced by inter-band transition in graphene and these
modes are undamped by the Landau-damping effect.

It is known that the elementary electronic excitation can be
achieved in an electron gas system through electronic transition
from occupied lower-energy states to the empty higher-energy states.
This includes both intra- and inter-band excitations. In a
semiconductor based C2DEG, because the band-gap between the
conduction band and the valence band is quite large, the inter-band
electronic transition requires a quite large excitation energy and
thus is less possible. However, graphene is a gapless 2DEG and the
inter-band electronic transition in graphene can therefore be
achieved via a small energy transfer. Hence, the elementary
electronic excitation via plasmon and coupled plasmon-phonon modes
induced by inter-band transition can be more possibly achieved in
graphene than in a C2DEG. Because the carrier density in graphene
can be effectively modulated up to $\sim 10^{13}$ cm$^{-2}$ by
applying the gate voltages \cite{jabl,liuy,bos}, the plasmon and
coupled plasmon-phonon excitations can be experimentally detected
and studied by different methods \cite{liuy,alt}. It is found
experimentally that the position of the G-peak induced by
optic-phonon scattering in the Raman spectrum in graphene varies
very little with varying the carrier density in the sample. However,
the Raman intensity around the G-peak depends quite strongly on the
carrier density in graphene. The results from the present study show
that for small values of $q$, optic-phonon$-$like modes can be
generated through coupled plasmon-phonon excitations via intra-band
(see $\Omega_+$ in Fig. 2 for $q<q_0$) and inter-band (see
$\Omega_1$ in Fig. 3) electronic transition channels. Such modes
depend rather weakly on $q$ and $n_e$ for small values of $q$. This
suggests that the position the optic-phonon$-$induced G-peak in the
Raman spectrum does not change significantly with varying the
carrier density in the sample, in line with the experimental
finding. From Fig. 2, we see that for relatively large values of
$q>q_0$ which decreases with increasing $n_e$, the frequency of
optic-phonon$-$like mode $\Omega_+$ induced by intra-band excitation
depends strongly on electron density. This implies that the Raman
intensity around the G-peak can be altered with varying the carrier
density in graphene. Because the optic-phonon$-$like mode induced by
inter-band excitation $\Omega_1$ depends very little on $n_e$ (see
Fig. 3), the change of the Raman intensity around the G-peak is
mainly induced by intra-band excitation mechanism. Thus, our results
can be applied to understand and explain the experimental findings.
These experimental and theoretical results indicate that the
electron-phonon interaction can affect strongly the features of
elementary electronic excitations in graphene. Furthermore, the
results shown in Figs. 2 and 3 demonstrate that the coupled
plasmon-phonon excitations in graphene can provide frequency-tunable
plasmonic modes. Hence, graphene can be used as plasmonic device for
various applications.

\section{Conclusions}

In this work, we have developed a tractable theory approach to study
the plasmon and coupled plasmon-phonon modes in graphene in which
the carrier-carrier interaction and the carrier-phonon coupling are
taken into consideration. We have also included the effect of the
presence of the inter-band electronic transition channels within our
calculations. The main conclusions obtained from this study are
summarized as follows.

An acoustic-like plasmon mode with a frequency $\omega_p\sim q^{1/2}
\sim n_e^{1/4}$ can be generated via intra-band electronic
transition in graphene and the carrier density dependence of this
mode differs from that in a conventional 2DEG. This is in line with
previous theoretical finding. We have found that an optical-like
plasmon mode with a frequency $\omega'_p$ can be excited via
inter-band electronic transition in graphene. However, this mode is
within the Landau-damping regime.

We have found that four coupled plasmon-phonon modes can be
generated in graphene. The features of two coupled-phonon modes
induced by intra-band transition channels differ significantly from
those observed in a conventional 2DEG system. Two optic-like modes
for coupled plasmon-phonon excitation can be observed via inter-band
electronic transition channels. Three coupled plasmon-phonon modes
depends strongly on electron density and plasmon wavevector.

The results obtained from this study demonstrate that if the
inter-band transition channels are taken into consideration, new
modes for plasmon and coupled plasmon-phonon excitations can be
obtained in graphene. The properties of the plasmon and coupled
plasmon-phonon excitations in graphene are very distinct from those
in the conventional 2DEG systems. Our results confirm that the
carrier-carrier interaction and the carrier-phonon coupling should
be equally considered to understand the properties of elementary
electronic excitation in graphene. Finally, our results presented
and discussed in this paper can be used to understand and explain
the experimental findings obtained from, e.g., the Raman
measurements.

\begin{acknowledgments}
Academy of Sciences, National Natural Science Foundation of China
and Department of Science and Technology of Yunnan Province, China.
One of us (H.M. Dong) was supported by Max Planck Society/Chinese
Academy of Sciences Doctoral Promotion Program.
\end{acknowledgments}


\begin{references}
\bibitem{ksn} K.S. Novoselov, A.K. Geim, S.V. Morozov, D. Jiang, Y. Zhang, S.V. Dubonos,
I.V. Grigoreva, and A.A. Firsov, Science 306 (2004) 666.

\bibitem{das} S. Das Sarma and E.H. Hwang, Phys. Rev. Lett. 102 (2009) 206412.

\bibitem{wxf} X.F. Wang and T. Chakraborty, Phys. Rev. B 75 (2007) 033408.

\bibitem{vra} V. Ryzhii, A. Satou, and T. Otsuji, J. Appl. Phys.
101 (2007) 024509.

\bibitem{fr} F. Rana, IEEE Transactions on Nanotechnology 7 (2008) 91.

\bibitem{jabl} M. Jablan, H. Buljan, and
Solja$\check{c}$i$\acute{c}$, Phys. Rev. B 80 (2009) 245435.

\bibitem{egm} E.G. Mishchenko, A.V. Shytov, and P.G. Silvestrov,
Phys. Rev. Lett. 104 (2010) 156806.

\bibitem{plaron} A. Bostwick, F. Speck, T. Seyller, K. Horn, M. Polini,
R. Asgari, A.H. MacDonald and E. Rotenberg, Science 328 (2010) 999.

\bibitem{pisa} S. Pisans, M. Lazzeri, C. Casiraghi, K.S. Novoselov,
A.K. Geim, A.C. Ferrari and F. Mauri, Nature Materials 6 (2007) 198.

\bibitem{liuy} Y. Liu and R.F. Willis, Phys. Rev. B 81 (2010) 081406(R).

\bibitem{bos} A. Bostwick, T. Ohta, T. Seyller, K. Horn, E. Rotenberg,
Nat. Phys. 3 (2007) 36.

\bibitem{pnas} D.A. Siegel, Cheol-Hwan Park, C.Y. Hwang,
Jack Deslippe, A.V. Fedorov, S.G. Louie and A. Lanzara, PNAS 108
(2011) 11365.

\bibitem{cst} C. Stampfer, F. Molitor, D. Graf, K. Ensslin, A. Jungen, C.
Hierold and L. Wirtz, Appl. Phys. Lett. 91 (2007) 241907.

\bibitem{dmb} D. M. Basko, S. Piscanec, and A. C. Ferrari, Phys. Rev. B
80 (2009) 165413.

\bibitem{moo} A. Mooradian and G.B. Wright, Phys. Rev. Lett.
16 (1966) 999.

\bibitem{tak} T. Iwanaga, T. Suzuki, S. Yagi and T. Motooka,
Appl. Phys. Lett. 86 (2005) 263102.

\bibitem{alt} G.C. Cho, T. Dekorsy, H.J. Bakker, R. H$\ddot{o}$v$\check{e}$l, and H. Kurz,
Phys. Rev. Lett. 77 (1996) 4062; H. Altan, X. Xin, D. Matten, and
R.R. Alfano, Appl. Phys. Lett. 89 (2006) 052110.

\bibitem{tse} W.K. Tse and S. Das Sarma, Phys. Rev. Lett. 99
(2007) 236802; T. Ando, J. Phys. Soc. Jpn. 76 (2007) 024712.

\bibitem{wxu} W. Xu, M.P. Das and L.B. Lin, J. Phys.: Condens. Matter
15 (2003) 3249.

\bibitem{pee} X.G. Wu, F.M. Peeters, J.T. Devreese, Phys. Rev.
B 32 (1985) 6982(R).

\end{references}
\end{document}